\documentclass[12pt]{article}
\input epsf.tex
\usepackage{graphicx}
\usepackage{psfig,picinpar,floatflt,amssymb,epsf,array}
\csname @addtoreset\endcsname{equation}{section}

\def\bseq{\begin{subequation}}  
\def\eseq{\end{subequation}}
\def\bsea{\begin{subeqnarray}}  
\def\esea{\end{subeqnarray}}

\newcommand{\beq}{\begin{equation}}
\newcommand{\eeq}{\end{equation}}
\newcommand{\bea}{\begin{eqnarray}}
\newcommand{\eea}{\end{eqnarray}}
\newcommand{\ena}{\end{eqnarray}}

\renewcommand{\a}{\alpha}
\renewcommand{\b}{\beta}

\renewcommand{\d}{\delta}

\newcommand{\Phib}{\bar{\Phi}}

\def\Mb{\kern 2pt\mathchoice
        {
         \vbox{\hrule width10pt height 0.4pt depth 0pt
         \kern 1.2pt\hbox{\kern -2pt$\displaystyle M$}}}
        {
         \vbox{\hrule width10pt height 0.4pt depth 0pt
         \kern 1.2pt\hbox{\kern -2pt$\textstyle M$}}}
        {
\vbox{\hrule width6pt height 0.4pt depth 0pt
         \kern 1.0pt\hbox{\kern -2pt$\scriptstyle M$}}}
        {
         \vbox{\hrule width5pt height 0.4pt depth 0pt
         \kern 0.8pt\hbox{\kern -2pt$\scriptscriptstyle M$}}}}

\def\Sb{\kern 2pt\mathchoice
        {
         \vbox{\hrule width6pt height 0.4pt depth 0pt
         \kern 1.2pt\hbox{\kern -2pt$\displaystyle S$}}}
        {
         \vbox{\hrule width6pt height 0.4pt depth 0pt
         \kern 1.2pt\hbox{\kern -2pt$\textstyle S$}}}
        {
         \vbox{\hrule width3.5pt height 0.4pt depth 0pt
         \kern 1.0pt\hbox{\kern -2pt$\scriptstyle S$}}}
        {
         \vbox{\hrule width3pt height 0.4pt depth 0pt
         \kern 0.8pt\hbox{\kern -2pt$\scriptscriptstyle S$}}}}

\def\Rb{\kern 2pt\mathchoice
        {
         \vbox{\hrule width5.5pt height 0.4pt depth 0pt
         \kern 1.2pt\hbox{\kern -2.5pt$\displaystyle R$}}}
        {
         \vbox{\hrule width5.5pt height 0.4pt depth 0pt
         \kern 1.2pt\hbox{\kern -2.5pt$\textstyle R$}}}
        {
         \vbox{\hrule width3.5pt height 0.4pt depth 0pt
         \kern 1.0pt\hbox{\kern -2.2pt$\scriptstyle R$}}}
        {
         \vbox{\hrule width3pt height 0.4pt depth 0pt
         \kern 0.8pt\hbox{\kern -2.2pt$\scriptscriptstyle R$}}}}

  \def\pp{{\mathchoice
      {
          \kern 1pt%
          \raise 1pt
          \vbox{\hrule width5pt height0.4pt depth0pt
            \kern -2pt
            \hbox{\kern 2.3pt
              \vrule width0.4pt height6pt depth0pt
              }
            \kern -2pt
            \hrule width5pt height0.4pt depth0pt}%
            \kern 1pt
       }
        {
          \kern 1pt%
          \raise 1pt
          \vbox{\hrule width4.3pt height0.4pt depth0pt
            \kern -1.8pt
            \hbox{\kern 1.95pt
              \vrule width0.4pt height5.4pt depth0pt
              }
            \kern -1.8pt
            \hrule width4.3pt height0.4pt depth0pt}%
            \kern 1pt
        }
        {
          \kern 0.5pt%
          \raise 1pt
          \vbox{\hrule width4.0pt height0.3pt depth0pt
            \kern -1.9pt  
            \hbox{\kern 1.85pt
              \vrule width0.3pt height5.7pt depth0pt
              }
            \kern -1.9pt
            \hrule width4.0pt height0.3pt depth0pt}%
            \kern 0.5pt
        }
        {
          \kern 0.5pt%
          \raise 1pt
          \vbox{\hrule width3.6pt height0.3pt depth0pt
            \kern -1.5pt
            \hbox{\kern 1.65pt
              \vrule width0.3pt height4.5pt depth0pt
              }
            \kern -1.5pt
            \hrule width3.6pt height0.3pt depth0pt}%
            \kern 0.5pt
        }
    }}

  \def\mm{{\mathchoice
               {
                 \kern 1pt
           \raise 1pt    \vbox{\hrule width5pt height0.4pt depth0pt
                  \kern 2pt
                  \hrule width5pt height0.4pt depth0pt}
                 \kern 1pt}
               {
                \kern 1pt
           \raise 1pt \vbox{\hrule width4.3pt height0.4pt depth0pt
                  \kern 1.8pt
                  \hrule width4.3pt height0.4pt depth0pt}
                 \kern 1pt}
               {
                \kern 0.5pt
           \raise 1pt
                \vbox{\hrule width4.0pt height0.3pt depth0pt
                  \kern 1.9pt
                  \hrule width4.0pt height0.3pt depth0pt}
                \kern 1pt}
               {
               \kern 0.5pt
         \raise 1pt  \vbox{\hrule width3.6pt height0.3pt depth0pt
                  \kern 1.5pt
                  \hrule width3.6pt height0.3pt depth0pt}
               \kern 0.5pt}
               }}

\def\pd{{\kern0.5pt
           + \kern-5.05pt \raise5.8pt\hbox{$\textstyle.$}\kern
0.5pt}}

\def\pmd{{\kern0.5pt
          \pm \kern-5.05pt
\raise6.3pt\hbox{$\textstyle.$}\kern1.5pt}}

\def\md{{\mathchoice
   {
      {{\kern 1pt - \kern-6.2pt \raise5pt\hbox{$\textstyle.$}\kern
1pt}}}
    {
      {{\kern 1pt - \kern-6.2pt \raise5pt\hbox{$\textstyle.$}\kern
1pt}}}
    {
      {\kern0.5pt - \kern-5.05pt
\raise3.4pt\hbox{$\textstyle.$}\kern0.5pt}}
    {
      {\kern0.5pt - \kern-5.05pt
\raise3.4pt\hbox{$\textstyle.$}\kern0.5pt}}}}

\begin{document}

\begin{titlepage}
{\hbox to\hsize{July  2003 \hfill
{Bicocca--FT--03--21}}}

\begin{center}
\vglue .06in
{\Large\bf Renormalizability of $\mathcal{N}=\frac{1}{2}$ Wess-Zumino model in \\
\vspace{5pt}
superspace}
\\[.45in]

Alberto Romagnoni\footnote{alberto.romagnoni@mib.infn.it}\\
{\it Dipartimento di Fisica dell'Universit\`a degli studi di
Milano-Bicocca,\\
and INFN, Sezione di Milano, piazza della Scienza 3, I-20126 Milano,
Italy}\\[.8in]

{\bf ABSTRACT}\\[.0015in]
\end{center}
In this letter we use the spurion field approach adopted in hep-th/0307099 in order to show that by adding $F$ and $F^2$ terms to the original lagrangian, the $\mathcal{N}=\frac{1}{2}$ Wess-Zumino model is renormalizable to all orders in perturbation theory. We reformulate in superspace language the proof given in the recent work hep-th/0307165 in terms of component fields.

${~~~}$ \newline

\end{titlepage}

It has been recently shown~\cite{seiberg, OV} that the IIB superstring
in the presence of a graviphoton background defines a superspace
geometry with nonanticommutative spinorial coordinates. This
deformation of superspace was previously considered
in~\cite{ferrara}. Field theories defined over $\mathcal{N}=1/2$
superspace (i.e.\ $\mathcal{N}=1$ euclidean superspace deformed by a
nonanticommutativity parameter $\{\theta^\a , \theta^\b \} = 2 C^{\a
  \b}$ with $C$ a nonzero constant) have been considered
in~\cite{seiberg,Rey,others,GPR,BF}.\\
\indent
In this non(anti)commutative superspace we study the Wess-Zumino model
\begin{equation}
S = \int d^8z \Phib \Phi - \frac{m}{2} \int d^6z \Phi^2
- \frac{\bar{m}}{2} \int d^6\bar{z} \Phib^2
- \frac{g}{3} \int d^6z \Phi \ast \Phi \ast \Phi - \frac{\bar{g}}{3} \int
d^6\bar{z} \Phib \ast \Phib \ast \Phib
\label{action}
\end{equation}
that in~\cite{seiberg} was shown to reduce to the usual WZ augmented
by a nonsupersymmetric component term $\frac{g}{6}\int d^4 x C^2 F^3 $
(with $C^2 = C^{\alpha \beta} C_{\alpha \beta} $).\\
\indent
In~\cite{GPR}, by introducing a spurion field~\cite{softbreaking}, $U
= C^2 \theta^2 \bar{\theta}^2$, to represent the supersymmetry
breaking term $F^3$, the divergence structure and renormalizability of
the $\mathcal{N}=1/2$ WZ model have been studied systematically in
superspace through two loops.\\
\indent
In this approach the classical action reads
\begin{equation} S =
\int d^8z \Phib \Phi - \frac{m}{2} \int d^6z \Phi^2 -
\frac{\bar{m}}{2} \int d^6\bar{z} \Phib^2 - \frac{g}{3} \int d^6z
\Phi^3 - \frac{\bar{g}}{3} \int d^6\bar{z} \Phib^3 + \frac{g}{6} \int
d^8z U (D^2 \Phi)^3\,.
\label{action2}
\end{equation}
\indent
It has been proven~\cite{GPR} that, up to this order, divergences are
at most logarithmic, that divergent terms have at most one
$U$-insertion (i.e.\ there is at most one power of $C^2$) and they are
of the form $F^{\alpha}\bar{G}^{k}$, with $\bar{G}= \bar{m} \bar{\phi}
+ \bar{g} \bar{\phi}^2$ and $\alpha \ge 1$, $\alpha + k \le 3$ (here
$\Phi| = \phi$, $D_\a \Phi| = \psi_\a$, $D^2 \Phi| = F$ and analogous
relations for the antichiral superfield). Finally, a counterterm of
the form $F^{\alpha}\bar{G}^{k}$ has been shown to be completely
equivalent to a counterterm of the form $F^{\alpha + k}$. After adding
by hand the terms $\int d^8z U (D^2 \Phi)^2$ and $\int d^8z U (D^2
\Phi)$, the model is renormalizable up to two loop order.\\
\indent
In the recent paper~\cite{BF} it has been shown that the same results
hold to all orders in perturbation theory: in particular the authors
of~\cite{BF}, working in terms of component fields, constrain the form
of divergent terms in the effective action using the two global
$U(1)$ (pseudo)symmetries of the theory~\cite{Rey} and making general
considerations on the structure and combinatorics of the Feynman
diagrams.\\
\indent
In this short letter we reformulate in superspace formalism the
discussion of~\cite{BF}, since this approach is usually more suitable
when some supersymmetry is left. We use the conventions
of~\cite{superspace}.

We parametrize the terms $F$ and $F^2$ in the classical lagrangian as
\begin{equation} \label{F-F2}
\lambda_1 g^3 \bar{m}^4 \int d^8z U (D^2 \Phi) + \lambda_2 g^2
\bar{m}^2 \int d^8z U (D^2 \Phi)^2 \,.
\end{equation}
We consider the two global $U(1)$ (pseudo)symmetries of the theory,
the $U(1)_{\Phi}$ flavor symmetry and $U(1)_{R}$
R-symmetry~\cite{Rey}.  In superspace language we have the charge
assignment given in table~\ref{tab1}.\\
\begin{table}[t]
\centering
\begin{tabular}{|c|c|c|c|c|c|c|c|}
\hline
  &  dim & $U(1)_{R}$ & $U(1)_{\Phi}$ &  & dim & $U(1)_{R}$ & $U(1)_{\Phi}$\\
\hline
$\Phi$ & 1 &   1        & 1 & $\bar{\Phi}$ &  1 & $-1$   & $-1$\\
\hline
$U$ & $-4$   & 4        & 0 & $d^4 \theta$ &  2  &  0   & 0\\
\hline
$D_{\alpha}$ & 1/2    &     $-1$   & 0 &
$\bar{D}_{\dot{\alpha}}$ &  1/2 & 1  & 0 \\
\hline
$D^2$ &   1 &     $-2$   & 0 &
$\bar{D}^2$ &   1 & 2  & 0 \\
\hline
$g$  &   0  &  $-1$  &  $-3$ &
$\bar{g}$ &  0  &  1  &  3 \\
\hline
$m$  &   1  &  0  &  $-2$ &
$\bar{m}$  &   1  &  0  &  2 \\
\hline
$\lambda_1$ &  0  &  0  &  0 &
$\lambda_2$ &  0  &  0  &  0 \\
\hline
\end{tabular}%
\caption{Global $U(1)$ charge assignment in superspace\label{tab1}}
\end{table}
\indent
In particular with the parametrization~(\ref{F-F2}) the coefficients
$\lambda_1$ and $\lambda_2$ are charge neutral under both $U(1)$
(pseudo)symmetries~\cite{BF}.\\
\indent
The most general divergent term in the effective action has the form
\begin{equation}
\int d^4 x \Gamma_{\mathcal{O}}= \lambda \int d^4 x d^4 \theta\,
(D^2)^{\gamma}\, (\bar{D}^2)^{\delta}\, (D_{\alpha} \partial^{\alpha
  \dot{\alpha}} \bar{D}_{\dot{\alpha}})^{\eta}\, \Box^{\zeta}\,
U^{\rho}\, \Phi^{\alpha}\, \bar{\Phi}^{\beta}
\end{equation}
with $\gamma$, $\delta$, $\eta$, $\zeta$, $\rho$, $\alpha$, $\beta$
non-negative integers. It is understood that every $D^2$, $\bar{D}^2$,
$D_{\alpha}$, $\bar{D}_{\dot{\alpha}}$, $\Box$, $\partial^{\alpha
  \dot{\alpha}}$ is acting on $U$, $\Phi$, $\bar{\Phi}$ superfields,
taking into account that
\begin{equation}
\begin{array}[b]{rclrcl}
D_{\alpha} \bar{\Phi} &=& 0\,,\qquad& 
\bar{D}_{\dot{\alpha}} \Phi &=& 0
\\[4pt]
\left[ D^{\alpha}, \bar{D}^2 \right] &=& i \partial^{\alpha
  \dot{\alpha}} \bar{D}_{\dot{\alpha}} \,,\qquad&
\left[ \bar{D}^{\dot{\alpha}}, D^2 \right] &=& i
\partial^{\dot{\alpha} \alpha} D_{\alpha}
\\[4pt]
D^2\bar{D}^2 D^2 &=& \Box D^2 \,,\qquad& 
\bar{D}^2 D^2 \bar{D}^2 &=& \Box \bar{D}^2\,.
\end{array}
\end{equation}
In our notation the coefficient $\lambda$, with dimension $d$ and
charges $q_{R}=R$ and $q_{\Phi}=S$, is
\begin{equation}
\lambda \thicksim \Lambda^d g^{x-R} \bar{g}^x \left( \frac{m}{\Lambda}
\right)^{y} \left( \frac{\bar{m}}{\Lambda} \right)^{y +
  \frac{S-3R}{2}} \lambda_2^{\omega_2}
\end{equation}
where $\Lambda$ is an ultraviolet momentum cutoff. $\lambda$ cannot be
a function of $\lambda_1$ since we cannot form a 1PI connected diagram
with a $\int U~(D^2 \Phi)$ term. Moreover $\omega_2 \le \rho$ since
$\lambda_2$ appears only in terms with a $U$ insertion
(see~(\ref{F-F2})).\\
\indent
Since the term $\Gamma_{\mathcal{O}}$ has dimension 4 and zero charge,
we have
\bea 
d &=& 2 + 4\rho - \alpha - \beta - \gamma - \delta -2 \eta -2 \zeta
\nonumber\\
R &=& \beta - \alpha + 2\gamma - 2 \delta - 4\rho 
\nonumber\\
S &=& \beta - \alpha\,.
\label{conditions}
\eea
The overall power of $\Lambda$ in $\Gamma_{\mathcal{O}}$ is
\begin{equation}
P = d -2y - \frac{S-3R}{2}
\end{equation}
and using eq.~(\ref{conditions})
\begin{equation}
P = 2 + 2\gamma - 2\rho - 2\alpha -4\delta - 2y - 2\eta - 2\zeta\,.
\end{equation}
Obviously we have a divergent contribution iff $P \ge 0$. We consider
the different cases:
\begin{itemize}

\item $\rho =0 $ 

It is the ordinary Wess-Zumino case.
\item $\rho =1 $ 

We have
\begin{equation} \label{r=1a}
\gamma - \alpha -2\delta - y - \eta - \zeta \ge 0\,.
\end{equation}
Since the $U$ superfield has only the $\theta^2 \bar{\theta}^2$
component, the $d^4 \theta$ integration acts on it. Moreover the
covariant $D^2$ derivatives can act only on $\Phi$ superfields (in
fact $D^3=0$, $D^2\bar{D}_{\dot{\alpha}} \bar{\Phi}=0$ and $D^2
\bar{D}^2 \bar{\Phi} = \Box \bar{\Phi}$), so we have
\begin{equation}
\gamma \le \alpha\,.
\end{equation}
Therefore the only possibility to satisfy~(\ref{r=1a}) is for
\begin{equation} \label{r=1b}
\gamma = \alpha,  \quad \delta=y=\eta=\zeta=0
\end{equation}
and we find the general divergent term
\begin{equation}
\label{diverg}
\int d^8z U (D^2 \Phi)^{\alpha} \bar{\Phi}^{\beta}\,.
\end{equation}
With the assignment~(\ref{r=1b}) we have $P=0$ showing that there is at most a logarithmic divergence.  

\item $\rho =1 + n$, $n > 0 $ 

Since the $U$ superfield has only the $\theta^2 \bar{\theta}^2$ component, we need at least $n~D^2$ and $n~\bar{D}^2$. Therefore
\begin{equation}
\gamma = n + \gamma_1\,, \qquad 
\delta = n + \delta_1
\end{equation}
and then
\begin{equation} \label{rho=1+n}
\gamma_1 - \alpha -2n -2\delta_1 - y - \eta - \zeta \ge 0\,.
\end{equation}
Since $\gamma_1 \le \alpha$ (as in the previous case), and $n > 0$, we
see that eq.~(\ref{rho=1+n}) cannot be satisfied.
\end{itemize}
\noindent 
In conclusion, we have only (logarithmic) divergent terms of the
form~(\ref{diverg}). Now we look for constraints on the coefficients
$\alpha$ and $\beta$ in order to show that there are only finitely
many divergent terms.\\
\indent
As seen in the discussion above, we have 
\begin{equation}
\rho=1\,,\qquad
\gamma=\alpha\,,\qquad
\delta=y=\eta=\zeta=0
\end{equation}
therefore $\lambda$ takes the form
\begin{equation}
\lambda \thicksim
g^{x-R}~\bar{g}^{x}~\bar{m}^\frac{S-3R}{2}~\lambda_2^{\omega_2}
\end{equation}
where
\begin{equation}
\frac{S-3R}{2} =  - \beta -2\alpha +6\,.
\end{equation}
If we look only at the UV divergent part of a diagram, the evaluation
of the integral cannot depend on the mass parameter (in fact in
dimensional regularization the divergences appear just as poles in
${1}/{\epsilon}$). Therefore powers of $\bar{m}$ in the coupling
constant $\lambda$ can appear:
\begin{itemize}
\item[-] from the vertex $\lambda_2 g^2 \bar{m}^2 \int d^8z U (D^2
  \Phi)^2$
\item[-] from the propagators $\langle \Phi \Phi \rangle = -
  \frac{\bar{m}~D^2}{p^2(p^2+m\bar{m})}~\delta^{(4)}(\theta-\theta')$.
\end{itemize}
Then, if we consider that the number of propagators $\langle \Phi \Phi
\rangle$ is always nonnegative and that $\omega_2 \le \rho$, we have
\begin{itemize}
\item $\omega_{2}=0 \quad \rightarrow \quad - \beta -2\alpha +6 \ge 0
  \quad\rightarrow \quad \beta + 2\alpha \le 6$
\item $\omega_{2}=1 \quad \rightarrow \quad - \beta -2\alpha +4 \ge 0
  \quad \rightarrow \quad \beta + 2\alpha \le 4$
\end{itemize}
We have also the condition $\alpha \ge 1$: in fact, after $D$-algebra
at least one $D^2$ survives and then, using~(\ref{r=1b}), there must
be at least one chiral $\Phi$ superfield.\\
\indent
To summarize, we have found that at any loop order the logarithmic
divergent terms have the form
\begin{equation}
\int d^8z U (D^2 \Phi)^{\alpha} \bar{\Phi}^{\beta} \,, \quad \alpha
\ge 1\,, 
\qquad
\beta + 2\alpha \le 6 - 2\omega_2\,,\quad\omega_2 =0, 1\,.
\end{equation}
\indent
Now we show that we can repackage them into the form
\begin{equation}
\int d^8z U (D^2 \Phi)^{\alpha} \bar{\mathcal{G}}^{k}
\end{equation}
with $\bar{\mathcal{G}}= \bar{m} \bar{\Phi} + \bar{g} \bar{\Phi}^2$
and $0 \le k \le 3 - \omega_2 - \alpha$.\\
\indent
The condition $y=0$ implies that in a divergent diagram the coupling
constant does not contain $m$ factors and so that there are not
propagators $\langle
\bar{\Phi} \bar{\Phi} \rangle$ (by the same observations done for
$\bar{m}$).  Therefore we have divergent contributions only from
diagrams without adjacent $\bar{\Phi}^3$ vertices. Then a divergent
diagram with $\bar{\Phi}$ external legs is analogous to a yet
divergent diagram with the insertion of $\bar{\Phi}^3$ vertices on
$\langle \Phi \Phi \rangle$ propagators. In fact, this operation does
not modify the divergence of the diagram, since $\langle \Phi \Phi
\rangle \thicksim \Lambda^{-4} \thicksim(\langle \Phi \bar{\Phi}
\rangle)^2$ and since the $D$-algebra is not modified if we look only
at divergent contributions. The only differences are the substitution
$\bar{m} \rightarrow \bar{g} \bar{\Phi}$ for every insertion and a
combinatorial factor ${q \choose k}2^{k}$ that takes into account the
${q\choose k}$ ways to insert $k$ vertices in $q$ $\langle \Phi \Phi
\rangle$ propagators and a symmetry factor $2=3\cdot \frac{1}{3} \cdot
2$ for every vertex.\\
\indent
Therefore, with this operation, it is possible to start with divergent
diagrams that give contributions to terms $\int U (D^2 \Phi)^{\alpha}$
at a given loop order, and build all possible diagrams that give
contributions to terms $\int U (D^2 \Phi)^{\alpha} \bar{\Phi}^{\beta}$
at the same order.\\
\indent
If we start with a divergent base diagram with fixed $\omega_2$ and
$\alpha$, and with symmetry factor $S$ (that we can understand to
include also the poles in ${1}/{\epsilon}$), the sum of all the
divergent contributions with $k \ge 1$ is
\begin{equation}
\label{sum}
S g^{x - \alpha +4} \bar{g}^{x} \lambda_2^{\omega_2} \bar{m}^{6 -
  2\alpha} \sum_{k=1}^{q} 2^{k} {q \choose k} \left(
\frac{\bar{g}}{\bar{m}} \right)^{k} \int d^8 z U(D^2 \Phi)^{\alpha}
\bar{\Phi}^{k}\,.
\end{equation}
Since
\bea
\sum_{k=1}^{q} 2^{k} {q \choose k} \left( \frac{\bar{g}}{\bar{m}}
\right)^{k}\bar{\Phi}^{k} &=& \left( 1 + 2\frac{\bar{g}}{\bar{m}}
\bar{\Phi} \right)^{q} - 1 
\nonumber \\
&=& \left( 1 + 4\frac{\bar{g}}{\bar{m}} \bar{\Phi} + 4
\frac{\bar{g}^2}{\bar{m}^2} \bar{\Phi}^2 \right)^{\frac{q}{2}} - 1 
\nonumber \\
&=&\left( 1 + 4\frac{\bar{g}}{\bar{m}^2} \bar{\mathcal{G}}
\right)^{\frac{q}{2}} - 1
\eea
and observing that for a diagram without $\bar{\Phi}$ external legs
($\beta=0$) $q = 6 - 2\omega_2 - 2\alpha$, we can finally rewrite
eq.~(\ref{sum}) as
\begin{equation}
S g^{x - \alpha +4} \bar{g}^{x} \lambda_2^{\omega_2} \bar{m}^{6 -
  2\alpha} \sum_{k=1}^{3 - \omega_2 - \alpha} 4^{k} {3 - \omega_2 -
  \alpha \choose k} \left( \frac{\bar{g}}{\bar{m}^2} \right)^{k} \int
d^8 z U(D^2 \Phi)^{\alpha} \bar{\mathcal{G}}^{k}
\end{equation}
wich agrees with the two loop results of~\cite{GPR}.\\
\indent
Taking into account that $\alpha=1,2,3$ we can conclude, in agreement
with~\cite{BF}, that to all orders in perturbation theory, the
divergent terms generated are (in component fields, with
$\bar{\mathcal{G}}|=\bar{G}$)
\bea
\omega_2=0 \qquad &\rightarrow& \qquad F, F^2, F^3, F\bar{G},
F^2\bar{G}, F\bar{G}^2 
\nonumber \\
\omega_2=1 \qquad &\rightarrow& \qquad F, F^2, F\bar{G}
\eea
\indent
Now we show that the counterterms $F, F^2, F^3$ are sufficient to
renormalize the theory. We can follow the argument of~\cite{GPR} to
claim that a contraction of any field with $\bar{G}$ is equivalent to
its contraction with $F$. This is possible also in completely
superspace language and translates into the equivalence
$\bar{\mathcal{G}} \rightarrow D^2 \Phi$. In fact, let us consider for
example the effect of a superfield factor $U(D^2 \Phi_{b})^2 [D^2 \Phi
  - \bar{m} \bar{\Phi}]$ as compared to $\bar{g}U(D^2 \Phi_{b})^2
\bar{\Phi}^2$ (here $\Phi_{b}$ is the background superfield). The
superfield propagators are
\bea
\langle \Phi \Phib \rangle &=& \frac{1}{ p^2 +
m \bar{m}} \d^{(4)}(\theta - \theta')
\nonumber\\
\langle \Phi \Phi \rangle &=& - \frac{\bar{m} D^2}{ p^2(p^2 + m
  \bar{m})} \d^{(4)}(\theta - \theta')
\nonumber\\
\langle \Phib \Phib \rangle &=& - \frac{m\bar{D}^2}{ p^2(p^2 +
  m\bar{m})} \d^{(4)}(\theta - \theta')
\eea
and from the Feynman rules for each chiral (antichiral) field there is
an extra $\bar{D}^2$ ($D^2$) derivative on each line leaving a vertex
(except for one of the lines at a (anti)chiral vertex).\\
\indent
In the Wick expansion, the operator $U(D^2 \Phi_{b})^2 [D^2 \Phi -
  \bar{m} \bar{\Phi}]$ can be contracted either with a $\Phi^3$
vertex, or with a $\bar{\Phi}^3$ vertex. Taking in account the
$D$-algebra (and in particular $D^2 \bar{D^2} D^2 = - p^2 D^2$), and
given the form of the propagators, in the first case the result is
zero. In the second case, the $D$-algebra is analogous and, given the
form of the propagators, the result is $\bar{g} U(D^2 \Phi_{b})^2
\bar{\Phi}^2$. In this last case there is a little subtlety: when we
contract $D^2 \Phi$ with $\bar{\Phi}$, after using $D^2 \bar{D^2} D^2
= - p^2 D^2$, we remain with a $D^2$ on the propagator $\langle\Phi
\bar{\Phi} \rangle$; this $D^2$ can be integrated by parts onto the
$\bar{\Phi}^3$ vertex, to give the exact Feynman rules for a vertex
$U(D^2\Phi_{b})^2 \bar{\Phi}^2$. We can treat in a similar way the
operators $U (D^2 \Phi_{b}) \bar{\mathcal{G}}$ and $U (D^2
\Phi_{b})\bar{\mathcal{G}}^2$, thus showing the equivalence of the two
forms of counterterms when inserted into diagrams.\\
\indent
Therefore the counterterms
\begin{equation}
\int U(D^2 \Phi), \int U(D^2 \Phi)^2, \int U(D^2 \Phi)^3 
\end{equation}
are sufficient to renormalize the theory at any order of perturbation
theory.

\vskip 30pt {\bf Acknowledgments}
I thank R.~Britto, B.~Feng, M.~Grisaru and S.~Penati for helpful
discussions. This work has been supported by INFN, MURST and the
European Commission RTN program HPRN-CT-2000-00131, in which I am
associated to the University of Padova.


\newpage
\begin{thebibliography}{100}

\bibitem{seiberg}
N.~Seiberg, \emph{Noncommutative superspace, $N=1/2$ supersymmetry,
  field theory and string theory}, JHEP {\bf 06} (2003) 010, 
hep-th/0305248.

\bibitem{OV}
J.~de~Boer, P.A. Grassi and P.~van Nieuwenhuizen,
\emph{Non-commutative superspace from string theory},
hep-th/0302078;\\
H.~Ooguri and C.~Vafa, \emph{The c-deformation of gluino and
  non-planar diagrams},\\ hep-th/0302109;\\
H.~Ooguri and C.~Vafa, \emph{Gravity induced c-deformation},
hep-th/0303063;\\
N.~Berkovits and N.~Seiberg, \emph{Superstrings in graviphoton
  background and $N=1/2 + 3/2$ supersymmetry}, JHEP {\bf 07} (2003) 010, 
hep-th/0306226.

\bibitem{ferrara}
S.~Ferrara and M.A. Lled\'o, \emph{Some aspects of deformations of
  supersymmetric field theories}, JHEP {\bf 05} (2000) 008, 
hep-th/0002084;\\
D.~Klemm, S.~Penati and L.~Tamassia, \emph{Non(anti)commutative
  superspace}, Class.Quant.Grav.
{\bf 20} (2003) 2905-2916, hep-th/0104190.

\bibitem{Rey}
R.~Britto, B.~Feng and S.-J. Rey, \emph{Deformed superspace, $N=1/2$
  supersymmetry and (non)renormalization theorems},
JHEP {\bf 07} (2003) 067, hep-th/0306215.

\bibitem{others}
S.~Terashima and J.-T. Yee, \emph{Comments on noncommutative
  superspace},\\ hep-th/0306237;\\
S.~Ferrara, M.A. Lled\'o and O.~Maci\'a, \emph{Supersymmetry in
  noncommutative superspaces}, JHEP {\bf 09} (2003) 068,
hep-th/0307039;\\
J.-H. Park, \emph{Superfield theories and dual supermatrix models},
JHEP {\bf 09} (2003) 046, hep-th/0307060;\\
T.~Araki, K.~Ito and A.~Ohtsuka, \emph{Supersymmetric gauge theories
  on noncommutative superspace}, hep-th/0307076;\\
R.~Britto, B.~Feng and S.-J. Rey, \emph{Non(anti)commutative
  superspace, UV/IR mixing and open Wilson lines},
JHEP {\bf 08} (2003) 001, hep-th/0307091.

\bibitem{GPR}
M.T. Grisaru, S.~Penati and A.~Romagnoni, \emph{Two-loop renormalization 
  for nonanticommutative $N=1/2$ supersymmetric WZ
  model}, JHEP {\bf 08} (2003) 003, \\ hep-th/0307099.

\bibitem{BF}
R.~Britto and B.~Feng, \emph{$N = 1/2$ Wess-Zumino model is
  renormalizable}, \\ hep-th/0307165.

\bibitem{softbreaking}
L. Girardello and M.T. Grisaru, Nucl. Phys. {\bf B194} (1982) 65.

\bibitem{superspace}
S.J. Gates Jr., M.T. Grisaru, M. Ro\v{c}ek and W. Siegel,
\emph{Superspace}, Benjamin Cummings, Reading 1983.

\end{thebibliography}
\end{document}